\documentclass[12pt,twoside]{article}
\usepackage{fleqn,espcrc1}

\usepackage{graphicx}
\usepackage[figuresright]{rotating}

\newcommand{\AmS}{{\protect\the\textfont2
  A\kern-.1667em\lower.5ex\hbox{M}\kern-.125emS}}

\title{Microscopic calculations for solar nuclear reactions\thanks{This 
work was partly supported by Grants D32513/FKFP-0242-2000/BO-00520-98 
(Hungary) and by the John Templeton Foundation (938-COS153).}}

\author{Attila Cs\'ot\'o\address{Department of Atomic Physics, E\"otv\"os 
University, \\ P\'azm\'any P\'eter s\'et\'any 1/A, H-1117 Budapest, Hungary}  
and Karlheinz Langanke\address{Institute for Physics and Astronomy, \\ 
Aarhus University, DK-8000 Aarhus, Denmark}} 

\begin{document}

\maketitle

\begin{abstract}
We have studied the $^4{\rm He}({^3{\rm He}},\gamma){^7{\rm Be}}$, 
$^3{\rm He}({^3{\rm He}},2p){^4{\rm He}}$, and 
$^7{\rm Be}(p,\gamma){^8{\rm B}}$ reactions of the solar p-p chain, using  
microscopic cluster models. Among other results, we showed that the $^6{\rm
Li}+p$ channel has a nontrivial effect on the $^7$Be-producing reaction, that
the existence of a resonance in $^6$Be close to the $^3{\rm He}+{^3{\rm He}}$
threshold is rather unlikely, and that the correlations between some
properties of $^7{\rm Be}/{^8{\rm B}}$  and the low-energy cross section of
$^7{\rm Be}(p,\gamma){^8{\rm B}}$ might help one to constrain the value of the
$S_{17}(0)$ astrophysical S-factor.
\end{abstract}

\section{INTRODUCTION}

The very-low-energy cross sections of the solar nuclear reactions are used as
input parameters in solar models \cite{Bahcall}. In order to get reliable
results from these models for the energy generation, neutrino fluxes, etc., of
our sun, all input parameters, including those coming from nuclear physics,
should be known as precisely as possible. Despite the tremendous improvements
in our understanding of these processes in the past decades, some
uncertainties still exist, which could considerably influence the predictions
of solar models \cite{Adelberger}. 

In order to try to reduce some of these uncertainties, we have performed a
systematic study of some of the key reactions of the solar p-p chain by using
a microscopic cluster model. The wave function of our model looks like 
\begin{equation}
\Psi=\sum_{L,S}{\cal A} \Bigg \{ \bigg [ \Big [\Phi^{A}
\Phi^{B} \Big ]_S\chi_L(\mbox{\boldmath $\rho$})
\bigg ]_{JM} \Bigg\},
\label{wf}
\end{equation}
where ${\cal A}$ is the intercluster antisymmetrizer,
$\Phi$ are the cluster internal states, $\mbox{\boldmath
$\rho$}$ is the intercluster relative coordinate, while $L$, $S$, and 
$J$ are the orbital angular momentum, intrinsic spin, and total spin,
respectively. The internal states of the $A\leq 4$ clusters are simple
$0s$ harmonic oscillator shell-model functions, while the heavier subsystems 
are described by two-cluster wave functions, similar to Eq.\ (\ref{wf}) (for
example, $^7$Be inside $^8$B is described as a $^4{\rm He}+{^3{\rm He}}$
two-cluster system). The intercluster relative motions $\chi$, which are the
most important degrees of freedom, are treated with high precision, using
variational methods \cite{meth}.

\section{SOLAR NUCLEAR REACTIONS}

We briefly summarize the physics motivation and the main results of
Refs.\ \cite{Be7,Be6,B8,B8other}, where the model, discussed above,
was applied to selected reactions of the solar p-p chain. Further details 
can be found in the original papers. 

${\it ^4{\it He}({^3{\it He}},\gamma){^7{\it Be}}}$: We studied the effects of
the $^6{\rm Li}+p$ channel on the reaction cross section and on the properties
of $^7$Be \cite{Be7}. It is known that the zero-energy astrophysical S-factor
($S(E)=\sigma (E)\exp [2\pi\eta (E)]$, where $\sigma (E)$ is the cross section
and $\eta$ is the Coulomb parameter) is correlated with the quadrupole moment
of $^7{\rm Li}/{^7{\rm Be}}$. This relation was used to estimate
the value of $S_{34}(0)$ \cite{Kajino}. As the $^6{\rm Li}+p$ configuration
brings in large charge polarization in $^7$Be, it might have some interesting
effects on these relations and on the determination of $S_{34}(0)$. We have 
shown that the inclusion of the $^6{\rm Li}+p/n$ channel in the description
of the $^4{\rm He}({^3{\rm He}},\gamma){^7{\rm Be}}$ and 
$^4{\rm He}({^3{\rm H}},\gamma){^7{\rm Li}}$ reactions really has an
appreciable effect \cite{Be7}.
However, detailed analyses indicate that in order to have a clear picture, one
should use a more sophisticated $^6$Li state than the one used by us ($\alpha
+d$ with pure $L=0$ relative motion between the $\alpha$ and the $d$). The
careful study of the correlations between the quadrupole moments of $^7{\rm
Be}/{^7{\rm Li}}$ and the reaction cross sections revealed that the currently
accepted $S_{34}(0)$ value seems to be in conflict with the recommended
quadrupole moment of $^7$Li.

${\it ^3{\it He}({^3{\it He}},2p){^4{\it He}}}$: We searched for narrow
resonances in $^6$Be, close to the $^3{\rm He}+{^3{\rm He}}$ threshold
\cite{Be6}. Such a state, if it exists, could strongly enhance the reaction
cross section and, as a consequence, suppress the fluxes of the $^7$Be- and
$^8$B-type solar neutrinos. We used methods which allowed us to rigorously
treat 2-body ($^3{\rm He}+{^3{\rm He}}$) or 3-body ($^4{\rm He}+p+p$)
resonances, if they exist. We found no indication for narrow resonances. It
seems to us that the only remaining possibility for the existence of such a
state is that it might come from the interplay between the $s$- and
$d$-states (not present in our model) of the two $^3$He clusters.

${\it ^7{\it Be}(p,\gamma){^8{\it B}}}$: This reaction is responsible for the
production of the highest-energy solar neutrinos (except for the
$hep$-neutrinos \cite{Bahcall}). Recently it has been the target of many
experimental and theoretical analyses. We would like to point out that a
possible way of finding the most probable values of the zero-energy
astrophysical S-factor, $S_{17}(0)$, of the reaction is to find correlations
between it and some measurable quantities of $^7$Be and $^8$B \cite{B8}. Fig.\
1 shows the results of our model for $S(0)$ in correlation with some
measurable properties of $^7{\rm Be}/{^8{\rm B}}$. We note that in Ref.\ 
\cite{B8}
an incorrect value was used for the phenomenological Coulomb-displacement
energy. We correct this error in Fig.\ 1.
\begin{figure}[htb]
\centering
\includegraphics[width=13.8cm]{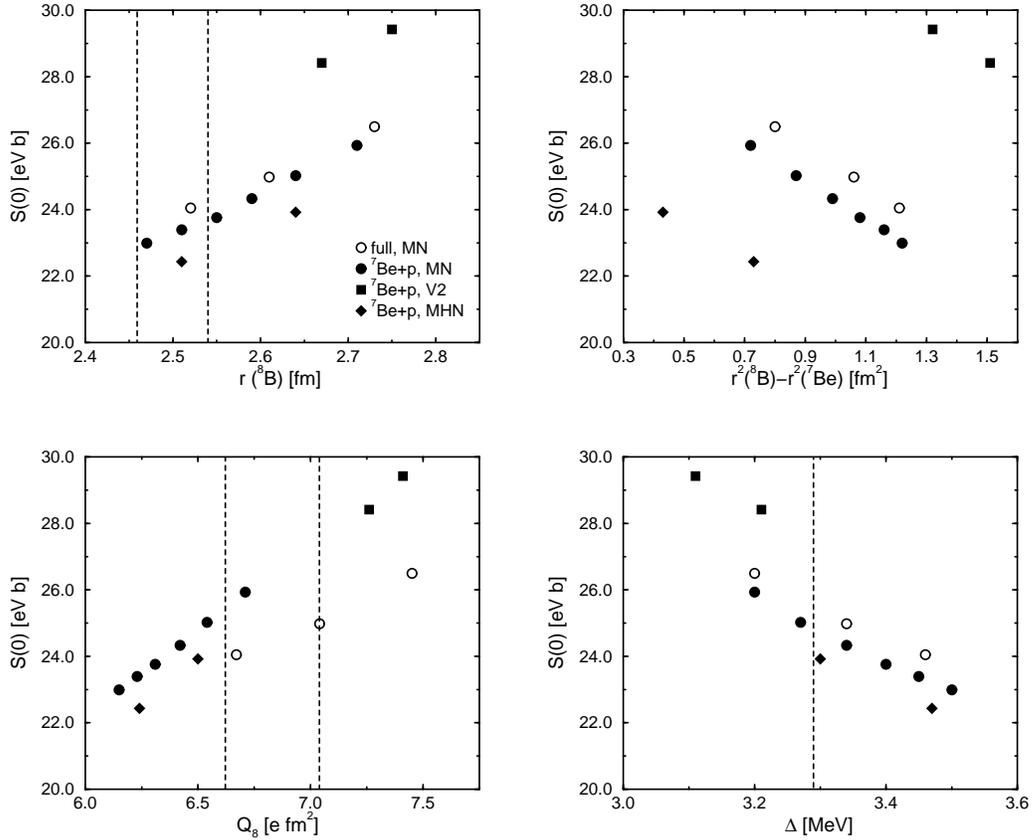}
\caption{Correlations between the zero-energy astrophysical
$S$ factor of the $^7{\rm Be}(p,\gamma){^8{\rm B}}$ reaction
and the $^8$B point-nucleon radius $r({^8{\rm
B}})$ (in fm), the $r^2({^8{\rm B}})-r^2({^7{\rm Be}})$
value (in fm$^2$), the $^8$B quadrupole moment $Q_8$
(in e$\,$fm$^2$), and the $\Delta=E({^8{\rm Li}})-E({^8{
\rm B}})$ Coulomb displacement energy (in MeV). The symbols show the results
coming from different N-N forces and model-spaces, while the vertical 
lines/bands indicate the phenomenologically suggested values/ranges. For 
the details, see Ref.\ \cite{B8}.}
\label{fig}
\end{figure}
As one can see, our model cannot produce a low value for
$S_{17}(0)$. It remains a question whether one could get a 
small $S_{17}(0)$ by going beyond the $^4{\rm He}+{^3{\rm He}}+p$ 
model-space.

\section{CONCLUSION}

We have studied some of the important nuclear reactions of the solar p-p
chain, using microscopic models. We believe that in order to make significant
further advances in the description of these processes, one should probably
use a full $A$-body dynamical model. However, the fully consistent description
of the bound and scattering states in such an approach will be a difficult
task.

\end{document}